\documentclass[fleqn,10pt]{wlscirep}
\usepackage[utf8]{inputenc}
\usepackage[T1]{fontenc}
\title{Conceptual winsorizing: An application to the social cost of carbon}

\author[1,2,3,4,5,6*]{Richard S.J. Tol}

\affil[1]{Department of Economics, University of Sussex, Falmer, BN1 9SL, United Kingdom}
\affil[2]{Institute for Environmental Studies, Vrije Universiteit, Amsterdam, The Netherlands}
\affil[3]{Department of Spatial Economics, Vrije Universiteit, Amsterdam, The Netherlands}
\affil[4]{Tinbergen Institute, Amsterdam, The Netherlands}
\affil[5]{CESifo, Munich, Germany}
\affil[6]{Payne Institute for Public Policy, Colorado School of Mines, Golden, CO, USA}

\affil[*]{r.tol@sussex.ac.uk}


\begin{abstract}
There are many published estimates of the social cost of carbon. Some are clear outliers, the result of poorly constrained models. Percentile winsorizing is an option, but I here propose conceptual winsorizing: The social cost of carbon is either a willingness to pay, which cannot exceed the ability to pay, or a proposed carbon tax, which cannot raise more revenue than all other taxes combined. Conceptual winsorizing successfully removes high outliers. It slackens as economies decarbonize, slowly without climate policy, faster with.
\end{abstract}
\begin{document}

\flushbottom
\maketitle
\thispagestyle{empty}

\section*{Introduction}
\begin{figure}
    \centering
    \includegraphics[width=1.0\linewidth]{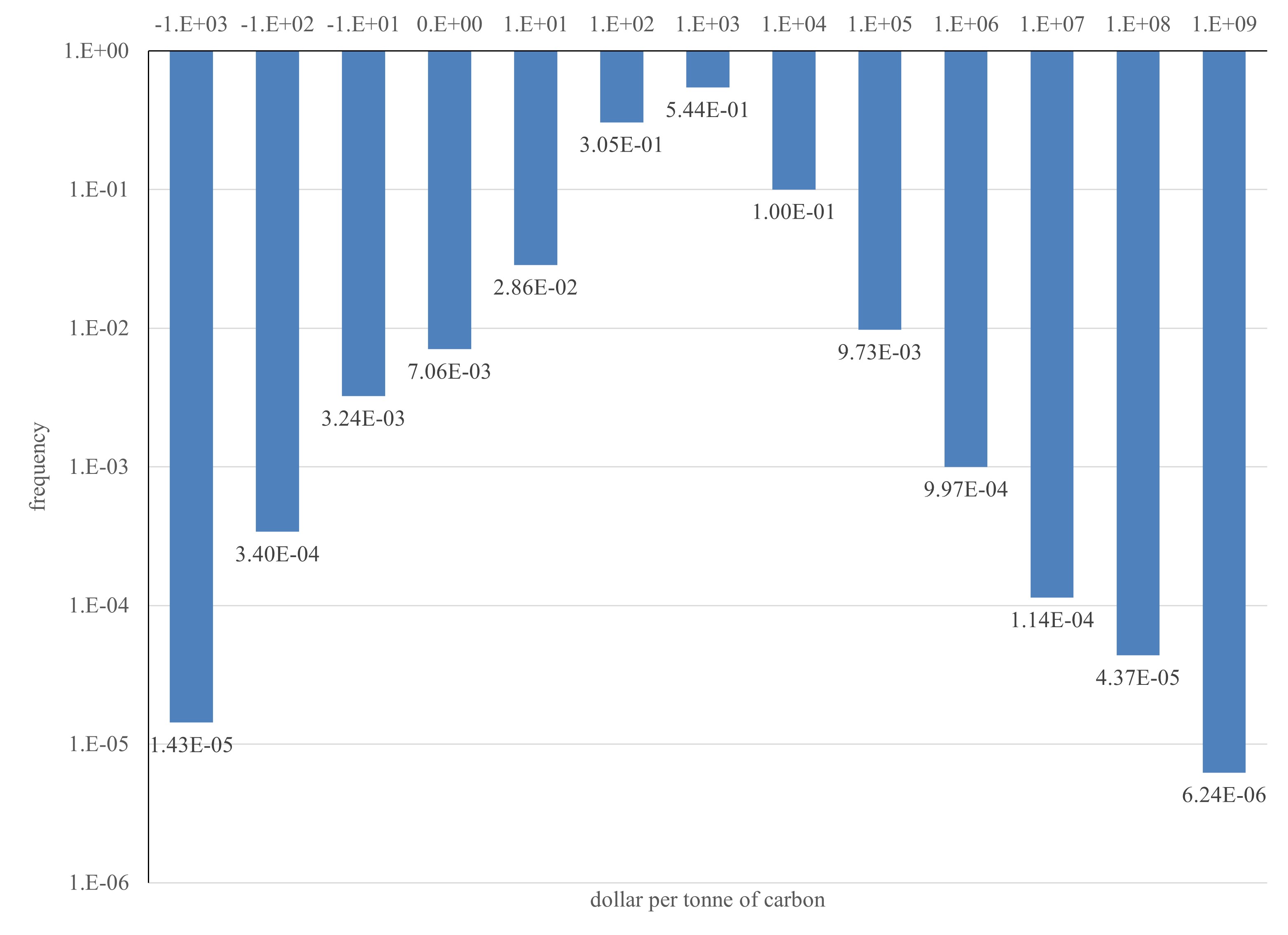}
    \caption{The histogram of published estimates of the social cost of carbon \footnotesize Both axes are on a logarithmic scale. The number displayed on the horizontal axis is the upper bound of the interval. For instance, 54.4\% of estimates fall between \$100/tC and \$1000/tC. Estimates are quality-weighted but not censored or winsorized.}
    \label{fig:hist}
\end{figure}

Winsorizing is a common way to reduce the impact of outliers. It replaces all observations $X_i$ above a threshold $T$ by that threshold. Estimates of the social cost of carbon range from -\$2,000/tC to +\$200,000,000/tC, spanning eleven orders of magnitude. See Figure \ref{fig:hist}. Clearly, some of these estimates are outliers. But which ones? Presumably, authors and referees applied some filter and there are unpublished estimates even more extreme. Should authors have censored more rigorously?

The social cost of carbon has multiple interpretations \cite{Kelleher2025}. I here focus on two. First, the social cost of carbon can be seen as the willingness to pay to reduce greenhouse gas emissions. This is constrained by the ability to pay.  In 2019, the carbon intensity of the world economy was 11,571 tC/\$. This implies a maximum willingness to pay of \$11,571/tC. All our money would go towards climate policy if we paid this amount.

Second, we can interpret the social cost of carbon as a proposed carbon tax. In 2019, governments collected 13.8\% of total income in taxation. A carbon tax of \$1,594/tC would allow for the abolition of all other taxes without a change in public spending. This is the Leviathan tax. If the carbon tax is larger, the public sector necessarily grows.

In an earlier meta-analysis\cite{Tol2023NCC}, I used these two numbers to censor or winsorize excessive estimates of the social cost of carbon. Other meta-analyses do not censor or winsorize.\cite{Bergh2014, Havranek2015, Wang2019, Moore2024PNAS, Yi2024} There is an implicit assumption in using the global average. Fewer than half of the people can meet the average ability to pay. If we use the global average as our anchor, we tacitly assume that richer people will pay for the emissions of poorer people. This may not be a realistic assumption.

\begin{figure}
    \centering
    \includegraphics[width=1.0\linewidth]{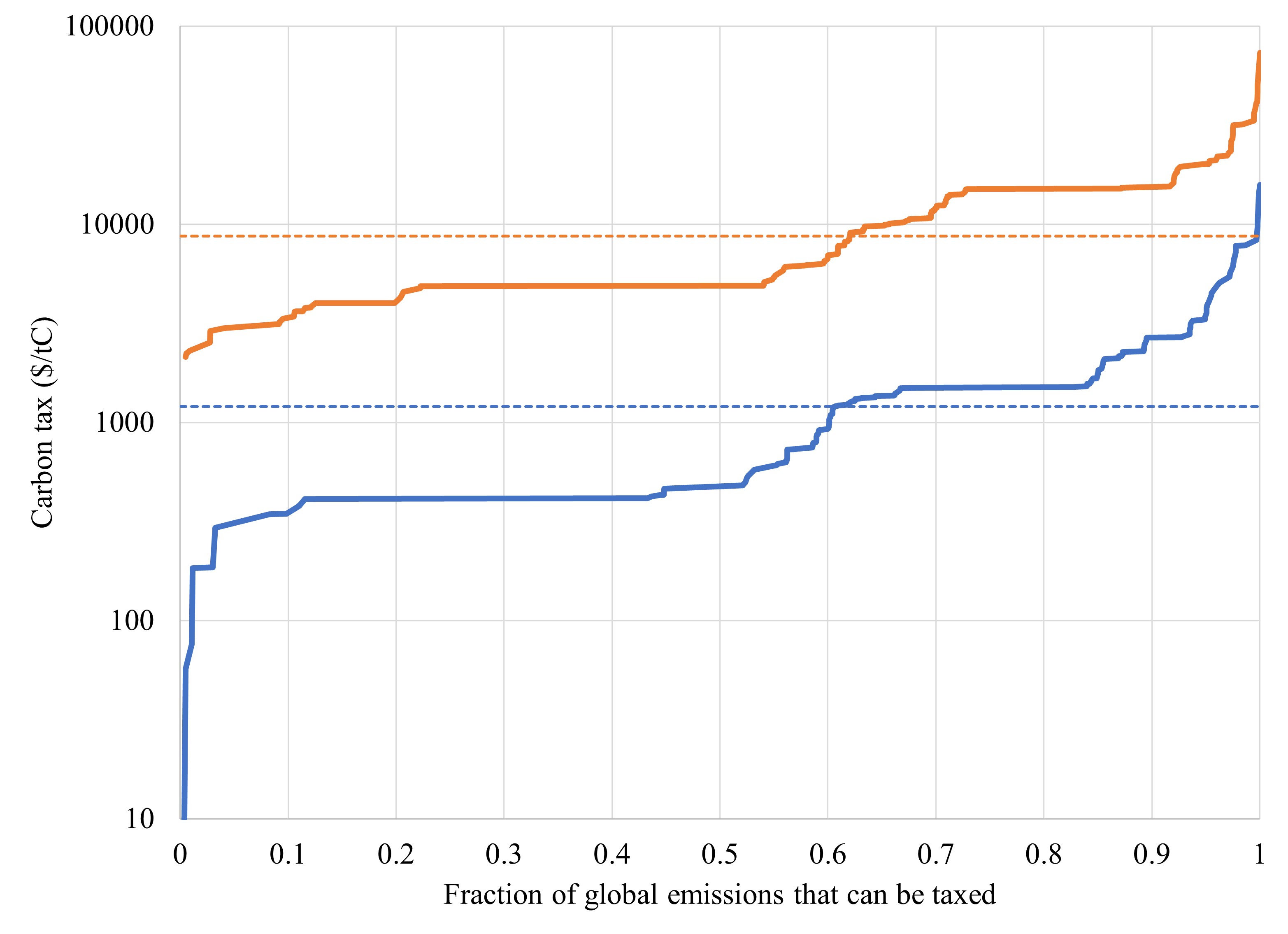}
    \caption{Limits on the social cost of carbon\footnotesize The maximum ability to pay (the carbon intensity of the economy) and the Leviathan tax (the carbon intensity times the tax share) in 2019 plotted against the share of global carbon dioxide emissions.}
    \label{fig:lev}
\end{figure}

Figure \ref{fig:lev} shows the maximum ability to pay by country. The lowest ability to pay is \$2,139/tC in Ukraine, the average \$8,743/tC. These numbers do not pose much of a limit on the published literature: 95\% (99\%) of published estimates fall below the minimum (average). Figure \ref{fig:lev} also shows the national Leviathan tax. These start at almost zero in Somalia and \$57/tC in Iraq. The average is \$1,205/tC. 92\% of social cost of carbon estimates fall below the average Leviathan tax, 19\% below Iraq's, and 1\% below Somalia's. The Leviathan tax does limit the social cost of carbon. It is \$415/tC for China, \$482/tC for India, and \$1,511 for the USA; 74\%, 76\%, and 93\% of published estimates.

Below, I use these data to winsorize published estimates of the social cost of carbon. I discuss how winsorizing may evolve over time with economic growth and climate policy before concluding.

\section*{Results}

\subsection*{Winsorizing}
\begin{figure}
    \centering
    \includegraphics[width=1.0\linewidth]{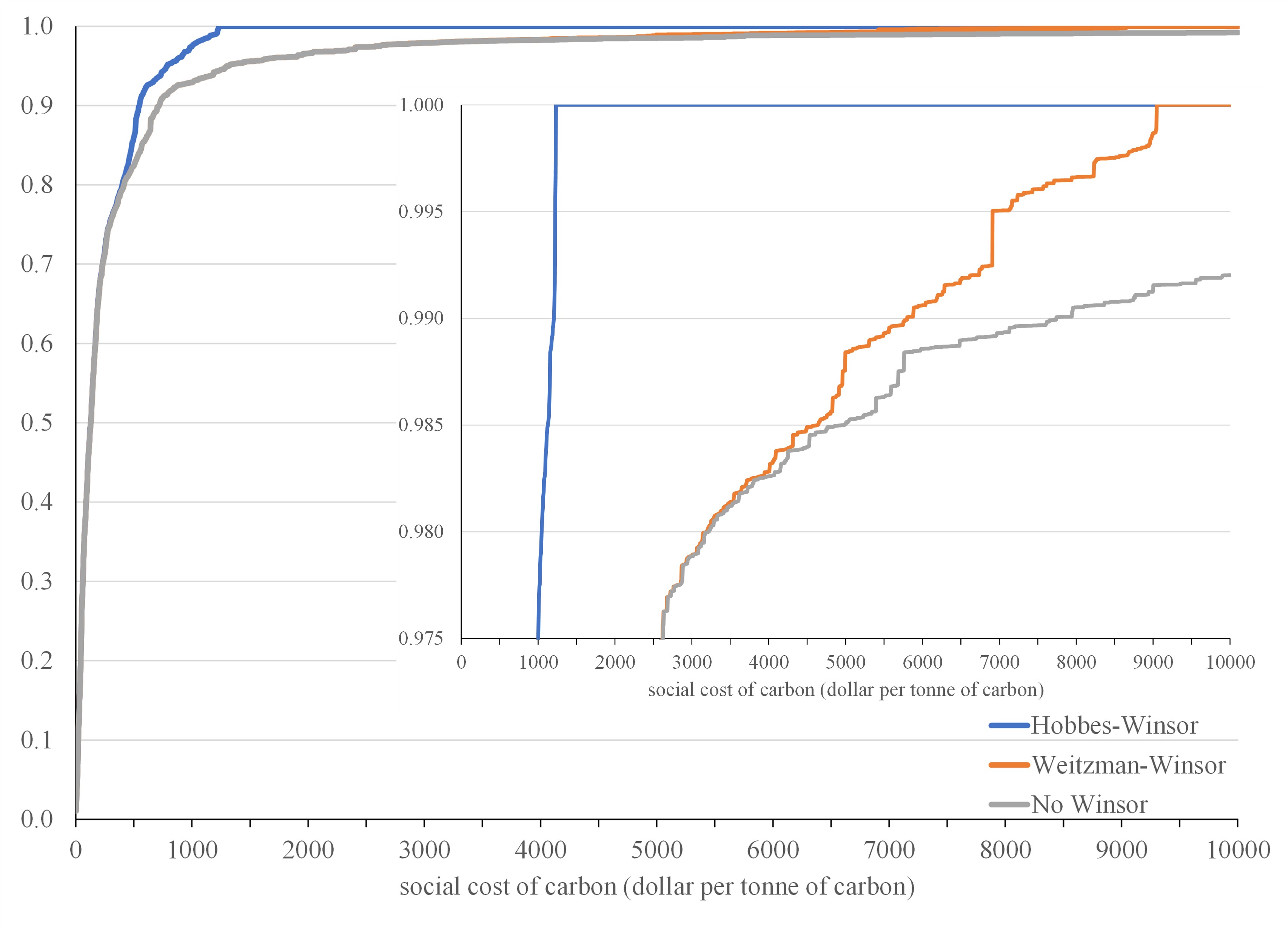}
    \caption{The cumulative density function without winsorizing, with winsorizing based on the ability to pay (Weitzman-Winsor), and with winsorizing based on the Leviathan tax (Hobbes-Winsor). The inset shows the same but with the horizontal axis restricted to show the tail.}
    \label{fig:cdf}
\end{figure}

Figure \ref{fig:cdf} shows the cumulative density function of the social cost of carbon, calculated in three ways. First, I use the social cost of carbon as reported. Estimates are weighted according to the prominence the original authors give to said estimates. Papers are weighted for quality.

Second, for each country, I winsorize the reported estimates with the national average ability to pay in 2019. I call this the Weitzman-Winsor, as paying this amount would bankrupt the country and drive utility to minus infinity.\cite{Weitzman2009} The global social cost of carbon is then the weighted average of the national, winsorized estimates, using the national share in global emissions as weights.

Third, for each country, I winsorize with the Leviathan tax\cite{Tol2012CCL} and average as above. I call this the Hobbes-Winsor, as paying more than this amount would necessarily increase the size of the public sector.\cite{Hobbes1651}

The Leviathan tax cuts off the tail at around \$1,200/tC. Ability to pay also cuts off the tail, but only at about \$9,000/tC. That may not seem like much of a tail-cut, but it has a dramatic effect on the moments of the distribution. See Table \ref{tab:stats}. In all three cases, the mode is \$18/tC and the median \$128/tC, emphasizing the profound right-skew in the distribution. Without winsorizing, the mean of the social cost of carbon is \$2,434/tC with a standard error of \$7,852/tC. The Weitzman-Winsor reduces this to \$375 (23)/tC, the Hobbes-Winsor to \$221 (6)/tC. The previously used censoring\cite{Tol2023NCC} reduces the mean social cost of carbon further, to \$196 (7)/tC.

\subsection*{Scenarios}
The proposed Weitzman-Winsor would become less stringent over time if the carbon intensity of the economy falls, as it is widely expected to do. The proposed Hobbes-Winsor further loses its bite with rising tax revenue in developing economies. As a result, the average social cost of carbon increases. On the other hand, the social cost of carbon increases over time as incremental emissions add to a worse problem.

The left panel of Figure \ref{fig:time} shows that the former effect dominates the latter. In all six alternative scenarios, the Hobbes-Winsor mean social cost of carbon increases by 80-120\% between 2020 and 2050.

\subsection*{Emission reduction}
Climate policy has three effects on top of the changes in baseline noted above: Emissions are lower but so is economic output. For moderate climate policy, the former falls faster than the latter, so the carbon intensity of the economy falls and the ability to pay and the Leviathan tax rise. Furthermore, the social cost of carbon falls as climate change slows. Together, this implies that winsorizing is less stringent if the social cost of carbon is imposed as a carbon tax\textemdash and this effect is stronger if the social cost of carbon is higher.

\begin{figure}
    \centering
    \includegraphics[width=0.49\linewidth]{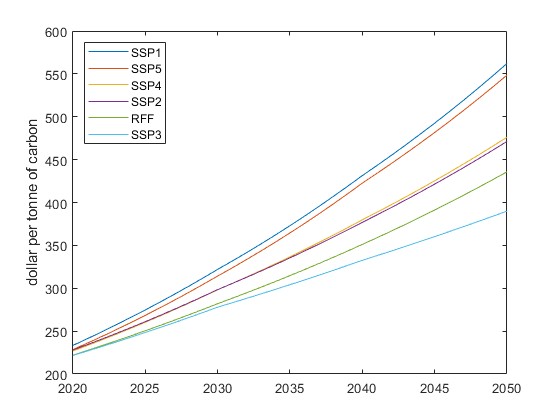}
    \includegraphics[width=0.49\linewidth]{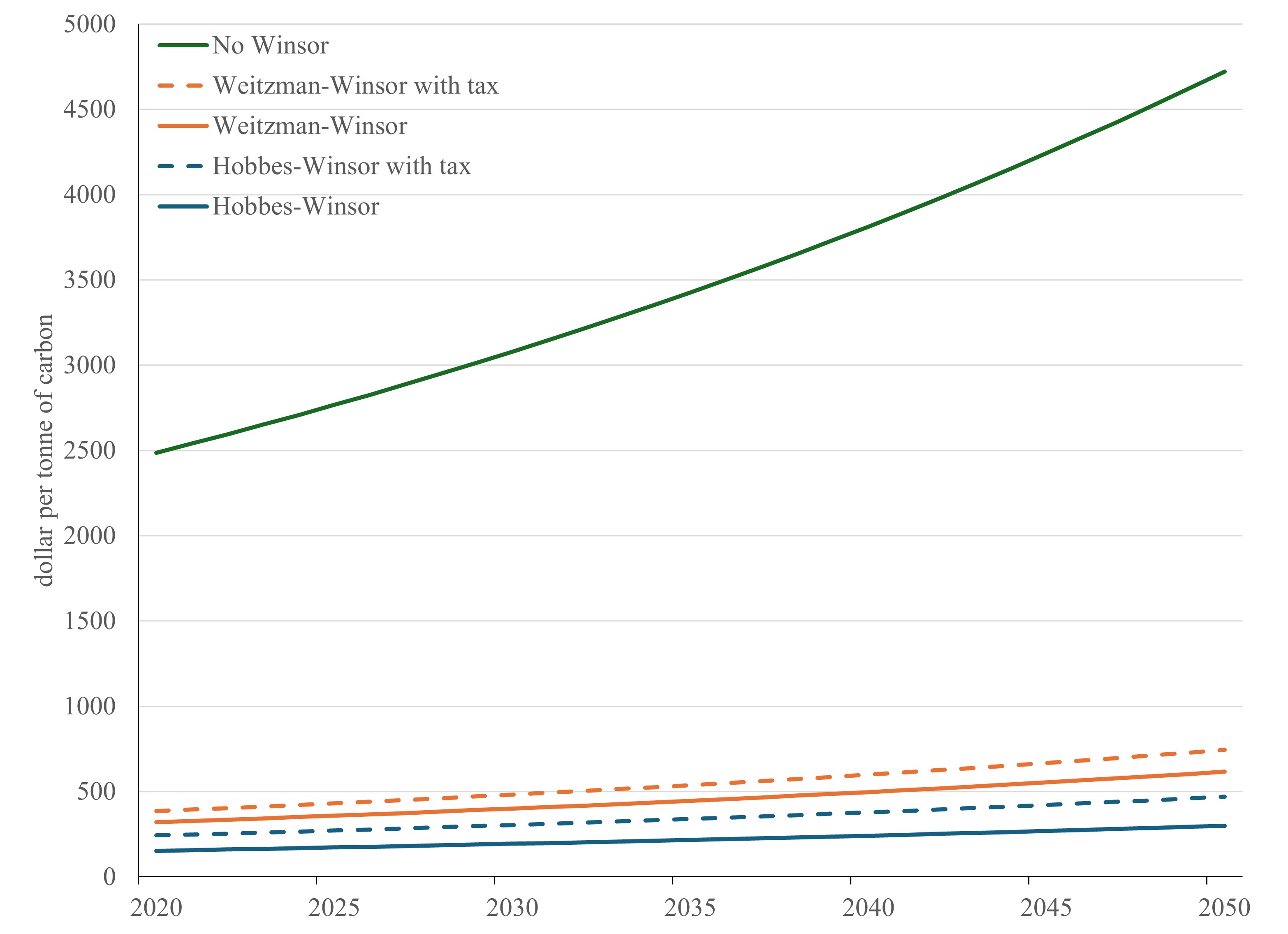}
    \caption{The mean of the social cost of carbon over time. Right panel: Hobbes-Windsor for 6 alternative scenarios. Left panel: Without Winsorizing and with the Weitzman- and Hobbes-Winsor with and without taxation.}
    \label{fig:time}
\end{figure}

The right panel of Figure \ref{fig:time} shows the effect. In all cases in the left panel, the social cost of carbon increases over time; it does so until the climate problem is solved.\cite{Ploeg2014IER} Figure \ref{fig:time} confirms Table \ref{tab:stats}: Winsorizing reduces the mean of the social cost of carbon, and more so with the Hobbes-Winsor than with the Weitzman-Winsor. As expected, taking the impact of a carbon tax on winsorizing into account considerably raises the mean of the social cost of carbon.

Without winsorizing, the mean social cost of carbon grows at the same annual rate (2.16\%) as the individual estimates. This increases to 2.20\% for the Weitzman-Winsor and to 2.27\% for the Hobbes-Winsor in the RFF scenario. In the SSP1 scenario, the growth rate increases to 2.97\% per year. Only in the SSP3 scenario (without tax) does the growth rate of the social cost of carbon outpace the rate of decarbonisation: The mean social cost of carbon increases by 1.90\% per year.

\section*{Discussion}
The social cost of carbon is a key parameter in the analysis of climate policy. It is uncertain so we should consider its mean,\cite{VNM1944} preferably from a wide range of studies to capture parametric, structural, and scenario uncertainty. The literature is wide-ranging, including outliers that have an unduly large impact on the sample mean. I here propose two variants of winsorizing, commonly used to deal with outliers. The first variant restricts, per country, the social cost of carbon to the ability to pay. The global mean social cost of carbon is the weighted average, using emissions as weights. This removes only a small part of the tail of the distribution of published estimates but considerably reduces the mean social cost of carbon. The second variants imposes a more arbitrary constraint: The public sector should not grow. This variant cuts off more of the tail of the distribution and further reduces the mean social cost of carbon. Over time, winsorizing becomes less severe, and the mean social cost of carbon increases. This would be more pronounced if the social cost of carbon were imposed as a carbon tax.

More granular data on income and energy use would refine the analysis but probably not change the qualitative result. The same holds for new or more detailed scenarios. Winsorizing here is limited to income and taxation, but one could also consider similar limits on the social cost of carbon due to constraints on labour and production. These would come to the fore with the social cost of methane: A tax on subsistence rice farming is out of the question. A carbon plus methane tax would hit coal miners particularly hard. Winsorizing based on population characteristics reduces the mean social cost of carbon on behalf of poor people's inability to pay. Reconciling this with equity weights, which have the opposite effect,\cite{Fankhauser1997} and compensation for adopting climate policy\cite{Harstad2007AER} is postponed to future research.

\section*{Methods}

\subsection*{Data}
Estimates of the social cost of carbon were collected over a number of years \cite{Tol2005, Tol2025anyas}. The current database \cite{Tol2025data} contains 14,152 estimates from 446 papers published between 1980 and 2024. There are 5,375 estimates from 264 papers for the growth rate of the social cost of carbon. The average, 2.01\% per year, is used to impute the social cost of carbon for 2019 from the originally reported year of emission. All estimates are expressed in 2015 U.S. dollars per metric tonne of carbon, using the producer price index.

Papers are quality-weighted on a scale from 1 to 4. The minimum score is 1. A paper scores one additional point if peer-reviewed, if it correctly computes the marginal, and if it uses a not-implausible scenario. Estimates are author-weighted, receiving higher weights if highlighted in the abstract or conclusion and lower weights if presented as a replication or sensitivity analysis.

\subsection*{Winsorizing}
The reported social cost of carbon $S_i$ is replaced by
\begin{equation}
    S_i^W = \frac{1}{\sum_c E_c} \sum_c E_c \min ( S_i, W_c )
\end{equation}
where $E_c$ are the carbon dioxide emissions in country $c$ and $W_c$ is its upper limit.

Without winsorizing, $W_c = \max S_i$. with the Weitzman-Winsor, $W_c = \frac{E_c}{Y_c}$, where $Y_c$ denotes GDP. With the Hobbes-Winsor, $W_c = \tau_c \frac{E_c}{Y_c}$, where $\tau_c$ is the fractional tax take.

\subsubsection{Scenarios}
For the growth rates of emissions and economic output, I follow the SSP scenarios\cite{RIAHI2017} as well as the RFF one.\cite{Rennert2022} I downscale the global scenarios to the country-level as follows. Poorer countries are assumed to grow faster: The national growth rate $g_{c.t}$ of country $c$ at time $t$ is $g_{c,t} = min(0.01, 0.059-0.005ln y_{c,t-1})$, where $y$ is per capita income. The coefficients are based on a regression of recent growth rate on income. National growth rates are rescaled so the global growth matches that in the SSP/RFF scenarios. Energy intensity of the economy and carbon intensity of the energy sector improve uniformly over space, follow the global SSP scenarios over time. Energy and carbon intensity in the RFF scenario follow the SSP2 scenario.

Observed tax revenue is updated with the growth rate of $0.026 + 0.016 ln y_{c,t-1})$. The parameters are from a regression of tax rates on per capita income.

For the social cost of carbon, I assume a growth rate of 2.16\% per year without a carbon tax and 1.95\% with \cite{Tol2023NCC}.

\subsubsection{Emission reduction}
Emissions would fall if the social cost of carbon is imposed as a carbon tax. This would reduce the emission intensity of the economy and so increase the ability to pay and Leviathan tax.

Assume a quadratic cost function:
\begin{equation}
\label{eq:cost}
    C_i = 0.5 \alpha_i R_i^2 Y_i  + \tau (1-R_i) E_i
\end{equation}
where $C_i$ are the costs of country $i$, $R_i$ is relative emission reduction, $E_i$ are uncontrolled emissions, $Y_i$ is output, $\alpha_i$ is a parameter and $\tau$ is the carbon tax. Cost-minimization implies
\begin{equation}
\label{eq:costeff}
    \frac{\partial C_i}{\partial R_i} = \alpha_i R_i Y_i -\tau E_i = 0 \Rightarrow R_i = \tau \frac{ E_i}{\alpha_i Y_i}
\end{equation}
At the global average, $R = 0.00126$ if $\tau = \$1/tC$ \cite{Tol2023EAP},  so $\alpha_i$ is chosen such that it is proportional to $\frac{E_i}{Y_i}$ and its emission-weighted global average equals $\frac{\bar{E}}{0.00126 \bar{Y}}$.

\bibliography{master}

\section*{Acknowledgements}
No funding supported this research.

\section*{Author contributions statement}
R.T. conceived of the study, collected the data, coded the algorithms, interpreted the results, and wrote the paper.

\section*{Additional information}
\subsection*{Code and data}
Code and data are on \href{https://github.com/rtol/metascc}{GitHub}.

\subsection*{Competing interests}
None.

\begin{table}[ht]
\centering
\begin{tabular}{|l|r|r|r|r|r|}
\hline
Winsor & mean & s.e. & s.d. & mode & median \\
\hline
None & 2,434 & 7,852 & 310,647 & 18 & 128 \\
\hline
Weitzman-Winsor & 375 & 23 & 926 & 18 & 128 \\
\hline
Hobbes-Winsor & 221 & 6 & 255 & 18 & 128 \\ \hline
Tol-censor & 196 & 7 & 249 & 18 & 128\\
\hline
\end{tabular}
\caption{\label{tab:stats}Descriptive statistics for the social cost of carbon (\$/tC) for three alternative winsorizings and previously used censoring\cite{Tol2023NCC}.}
\end{table}

\end{document}